\documentclass[11pt,oneside]{amsart}

\usepackage{amsmath}               % AmSLaTeX
\usepackage{times}
\usepackage{hyperref}

\numberwithin{equation}{section}

\title%
{Dynamics and Thermodynamics of Blackholes and Naked
Singularities II}

\author{Lorenzo Fatibene}
\address{Dipartimento di Matematica, Universit\`a di Torino, Italy}
\email{lorenzo.fatibene@unito.it}

\author{Mauro Francaviglia}
\address{Dipartimento di Matematica, Universit\`a di Torino, Italy}
\email{mauro.francaviglia@unito.it}

\author{Roberto Giamb\`o}
\address{Dipartimento di Matematica e Informatica,
Universit\`a di Camerino, Italy}
\email{roberto.giambo@unicam.it}
\urladdr{http://dmi.unicam.it/\~{}giambo}

\author{Giulio Magli}
\address{Dipartimento di Matematica, Politecnico di Milano, Italy}
\email{magli@mate.polimi.it}

\theoremstyle{plain}
\theoremstyle{plain}
\theoremstyle{plain}
\theoremstyle{plain}
\theoremstyle{definition}
\theoremstyle{remark}
\theoremstyle{definition}

\theoremstyle{plain}

\begin{document}

%\begin{abstract}

%\end{abstract}

\maketitle

\section{Introduction}\label{sec:intro}

The second edition of the international Workshop  ``Dynamics and Thermodynamics of
Blackholes and Naked Singularities`` took place at the Department
of Mathematics of  the Politecnico of Milano from May 10, 2007 to May 12, 2007.
Participation was restricted to 70 scientists due to
organizational reasons. The workshop was attended by people coming
from several different Countries. Sponsors of the conference were MIUR, SIGRAV
and INdAM-GNFM. As occurred for the previous edition held in 2003,
the workshop has been a fruitful occasion of
scientific exchange between people interested in various aspects
of blackhole theory, with special emphasis on Thermodynamics of Black Holes and Gravitational
Collapse. The main speakers, in alphabetical order, were:A. Burinskii, E. Elizalde, R. Giambo', L. Rezzolla, M. Cavaglia', L. Fatibene, P.S. Joshi, B. Schutz, C. Cercignani, S. Ferrara, B. Nolan. Chairmen of the contributed sessions were J. Kijowski (blackholes) and F. Giannoni (naked singularities).
The papers selected among those submitted for the Proceedings of the Conference
have now been e-published on the website of the Department
\href{http://www1.mate.polimi.it/bh2/}{http://www.mate.polimi.it/bh}, with a mirror copy
on the SIGRAV website. Each paper is available free of charge following the links given below.

\bigskip

\section{Contribution list}

\begin{enumerate}

\item{Stefano Ansoldi,\\ \emph{Spherical black holes with regular center}}\ \
\href{http://www1.mate.polimi.it/bh2/Ansoldi.pdf}{$\longrightarrow$}\\
\href{http://arxiv.org/abs/0802.0330}{arXiv:0802.0330v1 [gr-qc]}

\item{Mustapha Azreg-A\"{\i}nou,\\ \emph{Exact 5-dimensional cosmic string
solutions}}\ \
\href{http://www1.mate.polimi.it/bh2/Azreg.pdf}{$\longrightarrow$}

\item{David Brizuela, Jos\'e M. Mart\'in-Garc\'ia and Guillermo A. Mena Marug\'an, \\ \emph{Non-linear perturbations on spherical backgrounds (Formalism and algebraic implementation)}}\ \
\href{http://www1.mate.polimi.it/bh2/Brizuela.pdf}{$\longrightarrow$}

\item{Alexander Burinskii,\\ \emph{Black-holes, Regular Sources and Singularities: Kerr Schild Approach}}\ \
\href{http://www1.mate.polimi.it/bh2/Burinskii.pdf}{$\longrightarrow$}

\item{Marco Cavagli\`a,\\ \emph{Black Holes in particle colliders and in Earth's atmosphere}}\ \
\href{http://www1.mate.polimi.it/bh2/Cavaglia.pdf}{$\longrightarrow$}

\item{Saurya Das, Shanki Shankaranarayanan and Sourav Sur,\\ \emph{Power-law corrections to black-hole entropy via entanglement}}\ \
\href{http://www1.mate.polimi.it/bh2/Subramaniam.pdf}{$\longrightarrow$}\\ 
\href{http://arxiv.org/abs/0711.3164}{arXiv:0711.3164v1 [gr-qc]}

\item{Emilio Elizalde,\\ \emph{Large Quantum Vacuum Fluctuations at the Cosmological Level}}\ \
\href{http://www1.mate.polimi.it/bh2/Elizalde.pdf}{$\longrightarrow$}

\item{Lorenzo Fatibene, Marco Ferraris, Mauro Francaviglia, Gianluca Pacchiella,\\ 
\emph{Geometric Entropy for Self-Gravitating Systems in Alternative Theories of Gravity}}\\
\href{http://www1.mate.polimi.it/bh2/Fatibene.pdf}{$\longrightarrow$}

\item{Roberto Giamb\`o, Fabio Giannoni, Giulio Magli\\ \emph{Dynamics of homogeneous scalar fields with
self-interaction potentials}}\\
\href{http://www1.mate.polimi.it/bh2/Giambo.pdf}{$\longrightarrow$}\\
\href{http://arxiv.org/abs/math/0703512}{arXiv:math/0703512v1 [math.CA]}

\item{Jacek Jezierski, Jerzy Kijowski and Szymon \L\c{e}ski\\ \emph{Ground state of the two black holes system}}\ \
\href{http://www1.mate.polimi.it/bh2/Lenski.pdf}{$\longrightarrow$}

\item{Hiromi Saida,\\ \emph{Black Hole Evaporation as a Nonequilibrium Process}}\ \
\href{http://www1.mate.polimi.it/bh2/Saida.pdf}{$\longrightarrow$}\\
\href{http://arxiv.org/abs/0711.2330}{arXiv:0711.2330v2 [gr-qc]}

\item{Elizabeth Winstanley,\\ \emph{Hawking radiation from rotating brane black holes}}\ \
\href{http://www1.mate.polimi.it/bh2/Winstanley.pdf}{$\longrightarrow$}\\
\href{http://arxiv.org/abs/0708.2656}{arXiv:0708.2656v2 [hep-th]}

\end{enumerate}

\vspace{1truecm}

\textbf{The whole Proceedings volume is available at the 
\href{http://www2.mate.polimi.it/convegni/papers/acta_bh.pdf}{stable url}.}

\vspace{1truecm}

\end{document}